\documentclass[useAMS,usenatbib,referee]{mn2e}

\usepackage{graphicx}

\title[Galactic winds from Dwarf Galaxies]
{On the theory of mass loss in dwarf galaxies: I - basic equations and the case of wave/thermal driven winds}
\author[D. Falceta-Gon\c{c}alves]{D. Falceta-Gon\c{c}alves$^{1}$\thanks{E-mail:dfalceta@usp.br} 
\\$^{1}$Escola de Artes, 
Ci\^encias e Humanidades, 
Universidade de S\~ao Paulo, Rua Arlindo Bettio 1000, CEP 03828-000,
S\~ao Paulo, Brazil}

\begin{document}

\date{}

\pagerange{\pageref{firstpage}--\pageref{lastpage}} \pubyear{2012}

\maketitle

\label{firstpage}

\begin{abstract} 
 
In this work we present a semi-analytical model of galactic wind for dwarf galaxies where thermal and turbulent/momemtum driving mechanisms are studied. The model takes into account wave and internal adiabatic heating mechanisms, as well as radiative and adiabatic cooling. The importance of external sources of energy is discussed. We also studied the role of the spatial distribution of dark matter in the acceleration of the wind and on the mass loss rates. The basic model parameters that regulate the wind mass loss rate and terminal velocity are the gravitational potential profile, the equilibrium temperature of the gas and the amplitude of the turbulent motions of the gas. We found that dwarf galaxies are likely to present quasi-stationary winds {\bf with mass loss rates larger than $10^{-3}$M$_{\odot}$yr$^{-1}$ } even in the absence of turbulent motions (which is possibly related to the SNe feedback), if the interstellar gas is heated to $T > 10^4 - 10^5$K. We also found that the wind mass loss rate is enhanced for cusped dark matter distributions, such as the NFW-profile, due to the increased pressure gradients at the center of the galaxy. The solutions presented here may serve as benchmarks for numerical simulations, and as inputs for single zone chemical evolution models of dwarf galaxies.

\end{abstract}

      
\section{Introduction}

Outflows at galactic scales are ubiquitous in the Universe, typically related to the most energetic physical phenomena known, such as accretion into active galaxy nuclei (AGNs) and supernovae (SNe) driven winds (Veilleux, Cecil \& Bland-Hawthorn 2005, Chen et al. 2010, Steidel et al. 2010, Erb et al. 2012, Martin et al. 2012). Much less magnificent (and energetic) than these, but not less relevant, winds from less massive galaxies also play an important role on the chemical evolution of the intergalactic medium (IGM), and on the physical properties of the interstellar medium (ISM) of the galaxy itself (Dekel \& Silk 1986; Bradamante, Matteucci \& D'Ercole 1998; Fragile et al. 2003; Tremonti et al. 2004; Kirby, Martin \& Finlator 2011; Peeples \& Shankar 2011, and others).

Few tens of dwarf elliptical and spheroidal galaxies have been observed in the Local Group, most of them orbiting the Milky Way and M31. These are small objects, typically $< 1$kpc, but relatively massive, presenting dynamical masses of $M_* \sim 10^7 - 10^9$M$_{\odot}$, being a small fraction visible $M_* \sim 10^5 - 10^7$M$_{\odot}$ (Mateo 1998). The large mass-to-luminosity relation obtained indicates that these objects are dominated by dark matter (e.g. Armandroff, Olszewski \& Pryor 1995). 

Cosmological numerical simulations predict the existence of hundreds to thousands of low mass dark matter haloes orbiting large galaxies at groups as ours. In a possible scenario, most dark matter haloes may be associated to extremely low luminosity stellar components, which are simply not detected by current instruments (Somerville \& Primack 1999). In contrast to this, some numerical simulations predict a large number of merging and disruption of sattelites in the Local Group (e.g. Bullock \& Johnston 2005, Robertson et al. 2005). Galactic winds may remove the gas from dwarf galaxy haloes quenching star formation. The systematic low metallicity observed in dwarf galaxies also indicates that the star formation process is quenched at a given point of the galaxy evolution (Kirby, Martin \& Finlator 2011), possibly related to winds.

In the vast majority of galactic wind models star formation and SNe are assumed as main sources of energy and momentum (see e.g. Larson 1974; Chevalier \& Clegg 1985; Bradamante, Matteucci \& D'Ercole 1998; Stringer et al. 2012, Ruiz et al. 2013). Basically, these models assume that the SNe energy is thermalized at the ISM, increasing the gas thermal pressure that drives the galactic wind. The energy released by the SNe, or a fraction of it (typically $\sim 1 - 10\%$; Silich, Tenorio-Tagle \& Mu\~noz-Tu\~n\'on [2003]), may then be directly compared to the binding energy of the gravitational potential (e.g. Bradamante, Matteucci \& D'Ercole 1998; Lanfranchi \& Matteucci 2007, C\^ot\'e et al. 2012). This method is very simplified and do not account for different spatial distribution of the gas and the star forming regions, temperature distributions, and processes that may influence the dynamics of the wind, such as local turbulence and radiative losses. 

A complete description of the dynamics of SN ejecta interacting with the local ISM, and further acceleration of the galactic wind, is a very complex problem. Heating from cosmic rays, fractional ionization and radiative cooling are only few of the processes that make the effort of modeling the galactic winds even more difficult. Numerical simulations are able to describe many of these processes occurring in galactic outflows simultaneously (e.g. Recchi, Matteucci \& D'Ercole 2001; Mori, Ferrara, \& Madau 2002; Fragile et al. 2003; Fujita et al. 2004; Ruiz et al. 2012). However, different schemes and numerical resolutions provide  different results. It is well known that SPH schemes are unable to evolve shocks and certain instabilities, such as the Rayleigh-Taylor and Kelvin-Helmotz, properly. Even numerical recipes for injecting SNe energy in simulated galaxies are still under debate, from the randomness of starbursts in time and location, to the actual efficiency on the conversion of the SN thermal to ISM kinetic energy. For example, Recchi et al. (2001) simulate a single starburst with all stars at a narrow location, as well as Fujita et al. (2004), and assume that $3\%$ of the type II SNe energy feedback would thermalize with the ISM, which results in a single and gigantic hot bubble that drags all gas out of the galaxy. Mori et al. (2002), on the other hand, assumed a net net injection of energy into the ISM of $\sim 30\%$ of the SNe energy, but obtained similar results. Ruiz et al. (2012), on the other hand, showed that the buoyant instabilities may be responsible to up to $40\%$ of the mass loss of a dwarf galaxy. It is crucial therefore to provide the numerical community analytical (or semi-analytical) benchmarks.

In this work we present an analytical model of galactic winds in order to study the role of different dark matter density distributions, temperature distribution, and turbulent pressures, on the gas acceleration.  The model is applied for dwarf spheroidal galaxies. We particularly focus on the possibility of having strong winds, i.e. with mass loss timescales shorter than 1Gyr, in the absence of starbursts, for which thermally driven winds would suffice. We also study the scaling of the mass loss rate of the hybrid thermally/wave driven winds with the energy density of the kinetic perturbations, which may be correlated to star formation activity. The model is described in Section 2. In Section 3 we present the results where we compare different scenarios, followed by a brief discussion of their impact on the evolution of dwarf galaxies, in Section 4, and by the Conclusions.

\section{Basic equations}

If rotation does not play relevant roles in the dynamics of the interstellar gas we may assume radial symmetry for this problem, i.e. $\partial_\theta = \partial_\phi = 0$. The basic hydrodynamic equations to describe the gas are the mass continuity and momentum equations, given by:

\begin{equation}
\frac{\partial \rho}{\partial t} + {\bf \nabla} \cdot (\rho {\bf u}) = 0,
\end{equation}

\begin{equation}
\frac{\partial {\bf u}}{\partial t} + {\bf u} \cdot {\bf \nabla u} = -\frac{GM(r)}{r^2}-\frac{1}{\rho}\frac{\partial P}{\partial r}+{\bf g}_{\rm rad}+{\bf g}_{\rm wav},
\end{equation}

\noindent
where $u$, $P$ and $\rho$ represent the plasma velocity, thermal pressure and mass density, respectively. The additional terms ${\bf g}_{\rm rad}$ and ${\bf g}_{\rm wav}$ correspond to the specific net force generated by the radiation field and by waves, respectively. {\bf In this work we will also neglect the role of radiation pressure in the acceleration of the galactic wind, but will take into account the radiative losses in the energy equation. 

Typically, models of galactic winds assume the feedback from SNe as the main driver of such kinetic power. The injection of kinetic turbulence instead of the dynamical evolution of energy blasts when simulating SNe feedback have been discussed, e.g. by Falceta-Gon\c calves et al. (2010) and Scannapieco \& Bruggen (2010). Basically, the advantage of their recipe for SNe feedback is to bypass the thermal evolution of the SNe Sedov's phase. After this quasi-adiabatic expansion phase most of the energy is lost due to radiation, but part of the initial energy has been also converted into kinetic. The stochastic distribution of several SNe and star formation regions therefore would result in a turbulent medium. However, this prescription requires the adoption of an efficiency factor ($\eta$) for the conversion of the supernova energy into turbulence that is not easily estimated. As pointed by Ruiz et al. (2013), the generation of SN driven turbulence depends on the distribution of the blast waves in time and space. This because the ``snow plow'' phase stalls at scales of order of 10 - 100 pc, depending on the local ISM gas density and temperature. As an example, shown by their numerical simulations, the merging of hot cavities efficiently generate Rayleigh-Taylon unstable regions, while SNe localized far from the central region may brake the ISM and only a minor fraction of its energy is thermalized with the ISM. These effects result in a different scenario as the result of simultaneously and centrally localized packs of SNe.

In this paper, when solving the momentum and energy hydrodynamical equations, we will deal with the wave term as a kinetic pressure only. We do not attempt to relate these waves {\it  a priori} to any specific physical mechanism occuring in the galaxy, such as turbulence generated by gas infall or tidal forces, or by any internal source, such as supernovae and stellar winds.}

The acceleration provided by the wave pressure gradient is given by:

\begin{equation}
{\bf g}_{\rm wav} = -\frac{1}{2\rho} \frac{\partial \epsilon}{\partial r} \hat{r}, 
\end{equation}

\noindent
where $\epsilon$ represent the energy density of the waves, and is related to the averaged squared amplitude of the acoustic perturbations ($\left< \delta V^2 \right>$) as: 

\begin{equation}
\epsilon \equiv \rho \left< \delta V^2 \right>.
\end{equation}

The energy equation, given by:

\begin{equation}
\rho u \frac{\partial }{\partial t} \left( \frac{u^2}{2} +\frac{5 k_B T}{2 } - U_{\rm grav} \right) + \frac{u}{2} \frac{\partial \epsilon}{\partial r} = \frac{\epsilon (r)}{L}(u+c_s) - \Lambda n^2 + \Gamma,
\end{equation}

\noindent
where $\Lambda(\rm T)$ is the cooling function, $\Gamma$ the net heating source from external sources, $c_s$ the sound speed, $U_{\rm grav}$ the gravitational potential and $L(\rm r)$ the damping length of the acoustic waves, as a function of the radius.

The set of equations is closed by the differential equation that describes the gradient of the wave energy density. Under the WKB approximation, the energy density of the waves follows the relation:

\begin{equation}
\epsilon(r)=\epsilon_0 \frac{\mathcal{M}_0}{\mathcal{M}} \left( \frac{1+\mathcal{M}_0}{1+\mathcal{M}} \right)^2 \exp\left( - \int^r_{r_0} \frac{1}{L\left(r'\right)} dr'\right),
\end{equation}

\noindent
where we have defined the sonic Mach number $\mathcal{M} \equiv u/c_s$, and the index ``0" represents the variable value at $r=r_0$. Therefore, we may now obtain the differential form:

\begin{equation}
\frac{\partial \epsilon}{\partial r}=-\frac{\epsilon}{L}-\frac{\epsilon \left( 1+3\mathcal{M} \right)}{\left( 1+\mathcal{M} \right)} \left( \frac{1}{u} \frac{\partial u}{\partial r} - \frac{1}{T} \frac{\partial T}{\partial r} \right),
\end{equation}

Since we are mostly interested in the stationary solution of the hydrodynamic equations we will also consider $\partial_t = 0$. Now substituting Eqs. 1 and 7 in the momentum and energy equations (Eqs.\ 2 and 5), we obtain:

\begin{eqnarray}
u \frac{\partial u}{\partial r} = - \frac{GM}{r^2} - \frac{k_B T}{\mu m_{\rm H}}\left[ -\left( \frac{2}{r} +\frac{1}{u} \frac{\partial u}{\partial r} \right) + \frac{1}{T}\frac{\partial T}{\partial r} \right]+\nonumber \\ \frac{\epsilon}{2 \rho} \left[ \frac{1}{L} + \frac{\left(1+3\mathcal{M}\right)}{\left( 1+\mathcal{M} \right)} \frac{1}{u} \frac{\partial u}{\partial r} \right],
\end{eqnarray}

\noindent
and

\begin{eqnarray}
\frac{\partial T}{\partial r}=\frac{2}{3} \frac{T}{r} \left[ \frac{r \left( Q_w - n^2\Lambda + \Gamma \right)}{\rho u c^2_s} -\frac{r}{u} \frac{\partial u}{\partial r} -2 \right],
\end{eqnarray}

\noindent
respectively, with $Q_w = \epsilon(u+c_s)/L$ the total heating from the decaying energy of the waves.

The set of equations above can now be integrated spatially in order to give the radial distributions of the wind physical parameters.

\section{Results}

The physical processes that bring much complexity to the modeling of the galactic winds are related to the radiative heating and cooling of the plasma. Cosmic rays, ionizing radiation and the photoelectric effect on dust particles represent the main heat sources of the Galactic interstellar gas. These processes however are stricktly dependent on the stellar population itself, as well as on the dust properties (for the photoelectric effect) and magnetic field (for the diffusion of cosmic rays) of the ISM, making any attempt to quantify their contributions in early dwarf galaxies not practical analytically. 

In this sense, we will consider in this work two special cases. Firstly, we will assume the plasma to be isothermal. This is equivalent to assume that the timescales for radiative cooling and heating are much shorter than those of the wave heating and adiabatic cooling due to the expansion, and that the equilibrium between both radiative cooling and heating occurs at a given equilibrium temperature $T_{\rm eq}$. Secondly, we consider the full non-isothermal model, but assume that the external heating sources are irrelevant compared to the other terms of Equation 9, i.e. $\Gamma \rightarrow 0$. As discussed later, this is particularly interesting because even though not modeled properly, the heating sources ($\Gamma$) are shown to be extremely important for sustaining a steady galactic wind.

\subsection{The special case of isothermal galactic winds} 

In order of analytical simplicity let us start with the simplest case with no temperature gradients ($\partial T/\partial r = 0$), i.e. an isothermal galactic envelope. This assumption is not unrealistic in two conditions, when the strong UV radiation field of a starburst (or cosmic rays) is able to ionize most of the volume of the ISM of a dwarf galaxy, except for the clustered molecular clouds where star formation will continue to occur, or if the galaxy has no stars formed yet and the gas is basically molecular. In this case, the momentum equation (Eq.\ 8) is written in the form:

\begin{eqnarray}
\frac{1}{u} \frac{\partial u}{\partial r} \left[ u^2 - c_s^2-\frac{\left< \delta V^2 \right>}{2} \frac{1+3\mathcal{M}}{1+\mathcal{M}} \right]= \nonumber \\ -\frac{GM(r)}{r^2}+\frac{2 c_s^2}{r} + \frac{\left< \delta V^2 \right>}{2L},
\end{eqnarray}

\noindent
where the isothermal sound speed $c_s = (k_B T/\mu m_{\rm H})^{1/2}$ is used.

\subsubsection{for $\epsilon \rightarrow 0$}

Let us first consider the case where the energy density of acoustic waves is negligible compared to the thermal and gravitational energy densities. In this case we reduce the galactic wind problem to the standard Parker solution for a stellar wind:

\begin{eqnarray}
\frac{1}{u}\frac{\partial u}{\partial r} = \frac{\left( \frac{2c_s^2}{r} -\frac{GM(r)}{r^2} \right)}{\left( u^2 - c_s^2 \right)}.
\end{eqnarray}

Typically, for stars, the numerator at the fraction in the right of Eq. 11 is negative at the lower boundary of the wind. Thus this equation has an unique solution when both numerator and denominator of the right side fraction must vanish, at the critical radius. This position is the so-called sonic point, since $u(r_{\rm crit})=c_s$. The critical radius is obtained by making the numerator vanish.

Now, considering a galaxy instead, the position of $r_{\rm crit}$ depends on the mass of the galaxy $M(r_{\rm crit})$. The dark matter component dominates the gravitational potential of dwarf galaxies. 

The mass distribution of the dark matter halo may be described by a Navarro-Frenk-White (NFW) profile:

\begin{equation}
M^{\rm NFW}(r)=4\pi \rho_0 R_S^3\left[ \ln{\left( \frac{R_S+r}{R_S} \right)} - \frac{r}{R_S+r} \right],
\end{equation}

\noindent
where $R_S$ represents a characteristic radius and $\rho_0$ the core mass density of the dark matter. $M^{\rm NFW}$ diverges as $r \rightarrow \infty$ so we have to truncate the NFW profile at a given radius $R_{\rm trunc}$, which is assumed as 1kpc for this work. For larger radii the total enclosed mass of the galaxy will be constant, i.e. $M^{\rm NFW}(R_{\rm trunc})$. For the sake of simplicity we will only consider the dark matter mass contribution to the gravitational potential. This approximation is valid for all the cases we study in this paper, as the ratio of the dark matter to the gas potential is much larger than unity, for the radii computed ($r>10$pc). This approximation is important also because once the galactic wind is set the gas mass profile becomes time-dependent, in contradiction to the steady-wind approximation used in this work.
Similarly, we are not considering cosmological evolution of the halo dark matter mass or characteristic radius. Therefore we do not vary the characteristic radius, profile or total mass of the dark matter, as a function of redshift.

\begin{figure}
	\centering
		\includegraphics[scale=0.14]{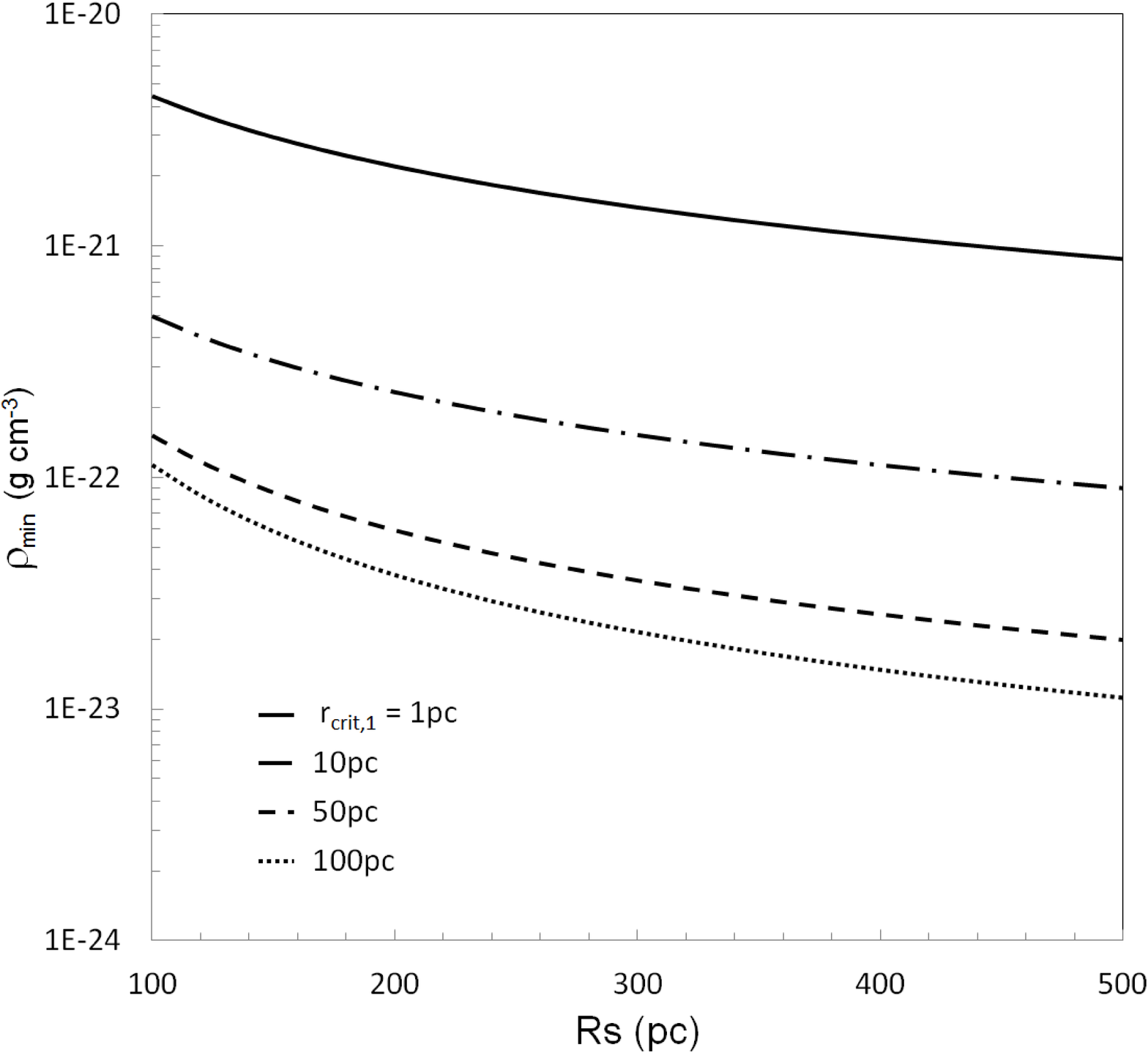}
		\caption {Minimal core mass density of the dark matter component needed for an accelerating wind at $r=r_{\rm crit,1}$, as a function of the characteristic radius $R_S$, obtained for $T=10^4$K. The relation has been calculated for $r_{\rm crit,1} = 1$pc (solid), 10pc (dash-dot), 50pc (dashed) and 100pc (dotted).}
	\label{fig1}
\end{figure}

At small radii $u \ll c_s$ and the gas is accelerated outwards only if the numerator of this equation is negative.
A major issue to the theory of pure isothermal winds in dwarf galaxies arises from the fact that the gravitational potential may be small at the central regions of the galaxy. As a consequence, the numerator of Eq. 11 will be positive resulting in $\partial_r u < 0$, i.e. the wind is actually a decelerating breeze. This will occur up to larger distances where the acceleration of gravity becomes dominant compared to the thermal component, and $\partial_r u$ becomes positive. We define the position where this transition occurs as the first critical point ($r_{\rm crit,1}$).

For a NFW dark matter distribution the critical location of the wind is obtained by solving the following equation, obtained by setting the numerator of Eq. 11 equals to zero: 

\begin{equation}
x=A \left[ \ln{\left( 1+x \right)} - \frac{x}{1+x} \right] ,
\end{equation}

\noindent
where $A=2\pi G\rho_0 R_S^3 c_s^{-2}$ and $x=r_{\rm crit}/R_S$. This equation may have up to 2 real solutions, as we discuss below. Depending on the gas temperature, if the core density $\rho_0$ is lower than a given threshold there is 0 or only 1 root for the above equation, which means that the gas is not thermally accelerated. The galactic wind will be stationary only in the cases where there are 2 real roots.

The first critical point depends on $T$, $\rho_0$ and $R_S$. In Figure 1 we show the dependence of the minimum density $\rho_0$ required, as a function of $R_S$, for the wind to be accelerated from $r_{\rm crit,1}$, for $T=10^4$K. For example, if the dark matter distribution has $\rho_0 = 10^{-21}$g cm$^{-3}$ and $R_S = 300$pc, the galactic wind starts to be accelerated at $r \sim 2$pc. Basically, at $r > r_{\rm crit,1}$ the thermal energy is converted into kinetic as long as the temperature is kept constant. 

Where the wind velocity reaches the sound speed the denominator of Eq. 11 vanishes, and a real solution for this problem is only obtained if the numerator vanishes as well. We define this position as the second critical point for the isothermal galactic wind $r_{\rm crit,2}$. This location represents the second root of Eq. 13.

The full solution of $u(r)$ is obtained by integrating Eq. 11 numerically, for which we used a 4$^{\rm th}$ order Runge-Kutta integration scheme. The calculations are performed from $r=10$pc, with a spatial resolution of $10^{-2}$pc, up to $r_{\rm max}=3000$pc. The mathematical solutions are independent on the minimum and maximum radii set, given that the sonic points are present in the chosen range. However, since energy injection should occur mostly at scales smaller than few tens of parsecs (e.g. starbursts) it is actually irrelevant to study the thermal acceleration at the most central regions of the galaxy. The gas temperature, $R_S$ and $\rho_0$ are initially set as constants. In order to obtain the critical solution at $r_{\rm crit,2}$ we varied the initial condition $u_0$.

\begin{figure}
	\centering
		\includegraphics[scale=0.13]{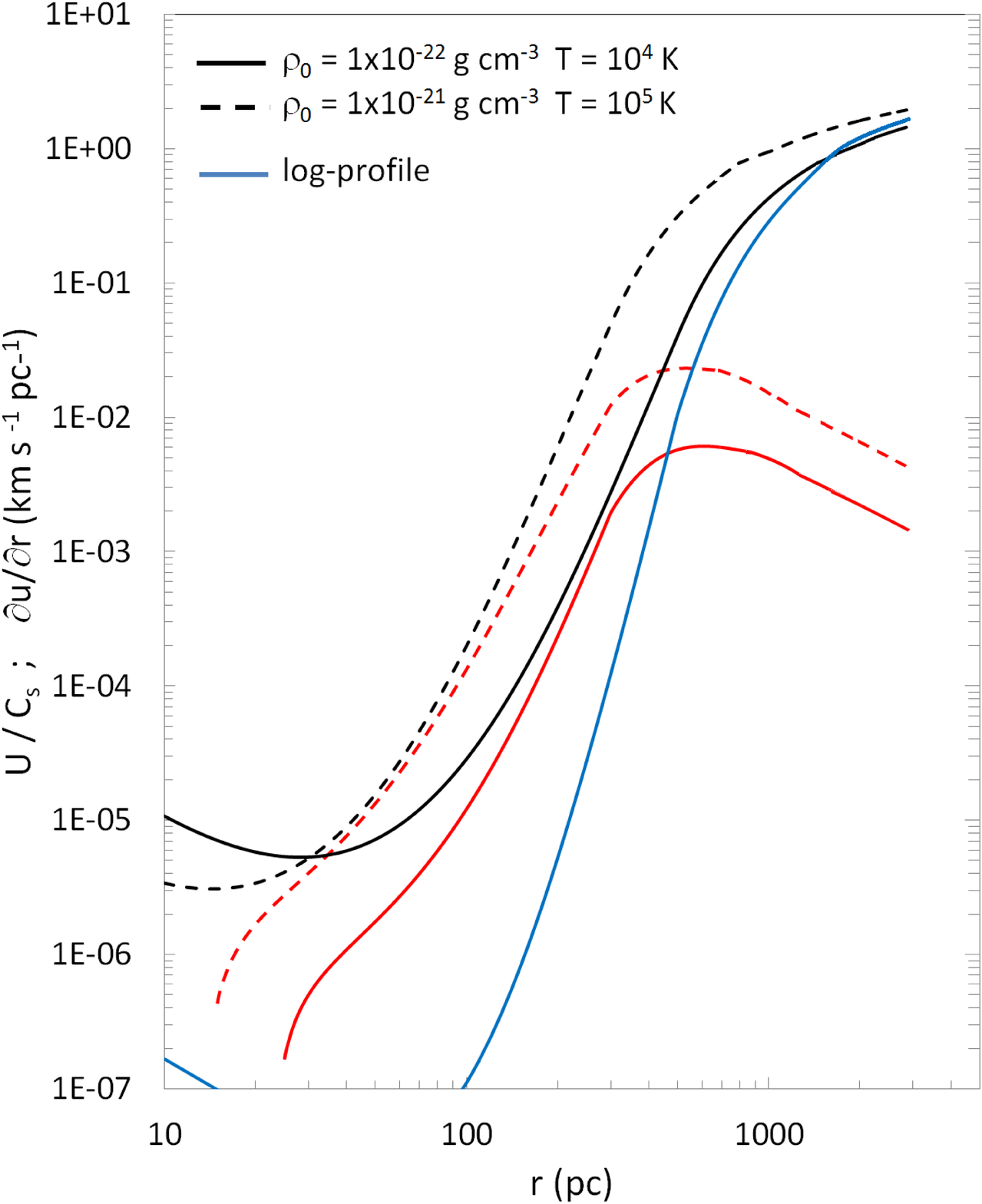}
		\caption {Wind sonic Mach number $M_s \equiv u/c_s$ and gradient of wind velocity $\partial_r u$ as a function of the radius, calculated for two models with $T=10^4$K and $\rho_0 = 10^{-22}$g cm$^{-3}$ (solid), and $T=10^5$K and $\rho_0 = 10^{-21}$g cm$^{-3}$ (dashed). $R_S = 200$pc for both cases. The third model shown represents the wind Mach number for a logarithmic profile of the dark matter density (see text).}
	\label{fig2}
\end{figure}

Two different models are presented in Figure 2 to illustrate the acceleration behavior of the thermal galactic wind. For these we set $R_S = 200$pc for both models, but $T=10^4$K and $\rho_0 = 10^{-22}$g cm$^{-3}$ (solid) and $T=10^5$K and $\rho_0 = 10^{-21}$g cm$^{-3}$ (dashed). We plot both the sonic Mach number ($u/c_s$) and the gradient of wind velocity $\partial_r u$ as a function of the radius. 

The galaxy is actually surrounded by the intergalactic medium, which eventually becomes dynamically important for the expansion of the wind. The stagnation of the wind occurs where the kinetic pressure of the wind is approximately balanced by the thermal pressure of the intergalactic medium, i.e. $\rho_{\rm wind}u_{\infty}^2 \sim n_{\rm IGM}kT_{\rm IGM}$. For $n_{\rm IGM} = 10^{-3}$cm$^{-3}$, $T=10^6$K and $u_{\infty} = 10 - 30$km s${-1}$, we obtain $\rho_{\rm wind} \sim 10^{-25} - 10^{-26}$g cm$^{-3}$, which corresponds to $r_{\rm stag} \sim 500 - 1000$pc. For the local Universe, such large IGM gas densities may occur at cluster of galaxies and extended gas haloes of massive galaxies, which the dwarfs may orbit. At high redshifts (e.g. $z>5$), deep gravitational potentials, such as clusters, do not exist yet and massive galaxies are probably just forming. In this case the wind of the low mass objects will basically interact with the diffuse IGM. The average density of baryons evolves with the scale factor $a$ as $n(a) ~ 3H_0^2\Omega_b (8 \pi G m_p a^3)^{-1} \sim 1.1 \times 10^{-5} (1+z)^3 \Omega_b h^2$cm$^{-3}$, and $n_{\rm IGM} = 10^{-3}$cm$^{-3}$ may be obtained for $z \simeq 9$. The asymptotic wind velocity obtained is typically few times larger than the sound speed, depending on the model. Therefore if the galactic wind stagnates at $500<r<1000$pc because of the pressure of the intergalactic medium (IGM) the wind velocity is not able to reach maximum values, and $u_{\infty} \sim c_s$ becomes a good approximation.

The gas mass loss rate is obtained by,

\begin{equation}
\dot{M}=4 \pi r^2 \rho_{\rm gas}(r) u(r),
\end{equation}

\noindent
which may be calculated at any position of the wind due to mass conservation, e.g. at $r=200$pc, where the models shown in Figure 2 result in $\dot{M} \simeq 3.9 \times 10^{-5} n_{200}{\rm (cm^{-3})}$ and $1.7 \times 10^{-3} \times n_{200}{\rm (cm^{-3})}$ M$_{\odot}$yr$^{-1}$ for $T=10^4$ and $10^5$K, respectively. As a result, after $\sim 1$Gyr, the total gas mass lost by an isothermal galactic wind of a typical dSph\footnote{assuming an initial gas mass of $10^7$M$_{\odot}$.} would be of $\sim 0.5$\% and $20$\% for the models with $T=10^4$ and $10^5$K, respectively. 

\subsubsection{The role of core or cusped dark matter profiles}

Recent observational and numerical works suggest that the NFW profile may not be the best description for the distribution of dark matter in dwarf galaxies (e.g. Burkert 1995; van den Bosch et al. 2000; de Blok \& Bosma 2002; Kleyna et al. 2003; Simon et al. 2005; Walker et al. 2009; Governato et al. 2010; Oh et al. 2011; Del Popolo 2012; Jardel \& Gebhardt 2012). Logarithmic profiles have been shown to match better the stellar velocity dispersion of the dSphs of the Local Group (see Persic, Salucci \& Stel 1996, Salucci et al. 2012). A different dark matter distribution can result in different accelerations of the galactic wind. In order to test this possible effect, and to compare to the standard NFW profile, we also consider a logarithmic gravitational potential for the dark matter density distribution which results in the following density and total enclosed mass distributions:

\begin{equation}
\rho_{\rm DM}\left( r \right) \simeq \frac{V_{\rm s}^2}{4\pi G} \frac{3r_{\rm s}^2+r^2}{\left(r^2+r_{\rm s}^2\right)^2},
\end{equation}

\begin{equation}
M^{\rm log}(r) \simeq \frac{V_s^2}{G} \frac{r^3}{r_s^2 + r^2},
\end{equation}

\noindent
where $V_s$ represents the virial circular orbital velocity at $r_s$, the characteristic radius for the dark matter distribution.

The advantage of using Eq. 16 is that the critical radii are obtained analytically. The first critical radius is given by:

\begin{equation}
r_{\rm crit}^{\rm log}=\left( \frac{2 c_s^2 r_s^2}{V_s^2-2c_s^2}\right)^{\frac{1}{2}} \simeq 45 {\rm \ pc},
\end{equation}

\noindent
considering $V_s = 100$km s$^{-1}$, $r_s = 100$pc and $T = 10^5$K. 

It is easy to notice that there is no second critical radius for the logarithmic profile. However, a saturation of the dark matter mass is expected to occur somewhere at $r > r_s$, mostly due to tidal disruption within the group of galaxies. If we define a cutoff radius for the dark matter distribution at $r=300 - 500$pc then the second critical radius occurs at $r_{\rm crit,1} \sim 1.4 - 2.9$kpc. 

For the logarithmic dark matter density profile the wind terminal velocity is similar to the one obtained with a NFW profile, as shown in Figure 2. For $T=10^5$K, we obtained asymptotic terminal velocities $u_\infty^{\rm NFW} \simeq 3.8c_s$ while $u_\infty^{\rm Log} \simeq 3.2c_s$, for a cutoff at $r=500$pc. The mass loss rate, on the other hand, is considerably different. We obtained $\dot{M} \simeq 2.1 \times 10^{-6} n_{200}{\rm (cm^{-3})}$ M$_{\odot}$yr$^{-1}$, which is 3 orders of magnitude lower than the obtained with a cusped NFW profile. The reason for this difference is that the wind is accelerated at larger radii for a log-profile, where the gas density is much lower.

\subsubsection{The case of $\epsilon \neq 0$ and $L \rightarrow \infty$}

SNe, stellar winds, tidal stirring and internal dynamic motions of the galaxy may drive turbulence in the interstellar medium. Once these turbulent waves are present the radial gradient of their kinetic pressure accounts for an extra acceleration term. This gradient is strongly dependent on the wave damping. If we assume that the waves propagate with no specific process of energy loss, except for the spatial dilution of the waves as they propagate outwards, i.e. $L \rightarrow \infty$, Equation 10 results in critical points which are exactly the same as in the case with no waves. However since the wind velocities now must be larger at the second critical point because of the extra term $\left< \delta V^2 \right>$, an increase in the mass loss rates and terminal wind velocities are expected. 

\begin{table*}
\begin{center}
\caption{Description of the models for isothermal winds}
\begin{tabular}{cccccccccc}
\hline\hline
Model & $\rho_0$(g cm$^{-3}$) & $R_S$(pc) & $V_s$(km s$^{-1}$) & $r_s$(pc) & $T$(K) & $\frac{\left< \delta V^2 \right>_0}{c_s^2}$ & $L$(pc) & $\frac{u_\infty}{c_s}$ & $\frac{\dot{M}}{n_{200}}$(M$_{\odot}$yr$^{-1}$cm$^{3}$) \\
\hline
IsoNFW1 & $10^{-22}$ & 200 & - & - & $10^4$ & 0.0 & $\infty$ & 2.1 & $3.9 \times 10^{-5}$ \\
IsoNFW2 & $10^{-21}$ & 200 & - & - & $10^5$ & 0.0 & $\infty$ & 3.8 & $1.7 \times 10^{-3}$ \\
\hline
IsoNFW3 & $10^{-21}$ & 200 & - & - & $10^5$ & 0.1 & $\infty$ & 4.2 & $2.2 \times 10^{-3}$ \\
IsoNFW4 & $10^{-21}$ & 200 & - & - & $10^5$ & 0.2 & $\infty$ & 4.1 & $3.0 \times 10^{-3}$ \\
IsoNFW5 & $10^{-21}$ & 200 & - & - & $10^5$ & 0.5 & $\infty$ & 4.3 & $6.4 \times 10^{-3}$ \\
IsoNFW6 & $10^{-21}$ & 200 & - & - & $10^5$ & 1.0 & $\infty$ & 4.4 & $1.7 \times 10^{-2}$ \\
\hline
IsoNFW7 & $10^{-21}$ & 200 & - & - & $10^5$ & 0.5 & $20$ & 3.4 & $1.6 \times 10^{-3}$ \\
IsoNFW8 & $10^{-21}$ & 200 & - & - & $10^5$ & 0.5 & $50$ & 3.5 & $1.7 \times 10^{-3}$ \\
IsoNFW7 & $10^{-21}$ & 200 & - & - & $10^5$ & 10.0 & $20$ & 3.4 & $1.6 \times 10^{-3}$ \\
IsoNFW8 & $10^{-21}$ & 200 & - & - & $10^5$ & 10.0 & $50$ & 3.6 & $1.9 \times 10^{-3}$ \\
IsoNFW9 & $10^{-21}$ & 200 & - & - & $10^5$ & 10.0 & $100$ & 3.5 & $6.1 \times 10^{-3}$ \\
\hline
IsoLog1 & - & - & 100 & 100 & $10^5$ & 0.0 & $\infty$ & 3.2 & $2.1 \times 10^{-6}$ \\
IsoLog2 & - & - & 100 & 200 & $10^5$ & 0.0 & $\infty$ & 2.4 & $6.0 \times 10^{-6}$ \\
\hline
IsoLog3 & - & - & 100 & 200 & $10^5$ & 0.2 & $\infty$ & 2.6 & $1.5 \times 10^{-5}$ \\
IsoLog4 & - & - & 100 & 200 & $10^5$ & 0.5 & $\infty$ & 2.7 & $5.9 \times 10^{-5}$ \\
IsoLog5 & - & - & 100 & 200 & $10^5$ & 1.0 & $\infty$ & 2.8 & $6.4 \times 10^{-4}$ \\
\hline
IsoLog6 & - & - & 100 & 200 & $10^5$ & 1.0 & $20$ & 3.2 & $7.4 \times 10^{-4}$ \\
IsoLog7 & - & - & 100 & 200 & $10^5$ & 10.0 & $20$ & 3.3 & $7.1 \times 10^{-4}$ \\
IsoLog8 & - & - & 100 & 200 & $10^5$ & 10.0 & $100$ & 3.3 & $2.5 \times 10^{-3}$ \\
\hline\hline
\end{tabular}
\end{center}
\end{table*}

In Figure 3 we present the wind velocity profiles considering $T=10^5$K and a NFW profile with $\rho_0 = 10^{-21}$g cm$^{-3}$ (dashed) and $R_S = 200$pc, for wave amplitudes $\left< \delta V^2 \right>_0 = 0$ (dotted), $0.1 c_s^2$ (dot-dashed), $0.2 c_s^2$ (dashed) and $0.5 c_s^2$ (solid). The wind terminal speeds and mass loss rates are also shown in Table 1. 

\begin{figure}
	\centering
		\includegraphics[scale=0.13]{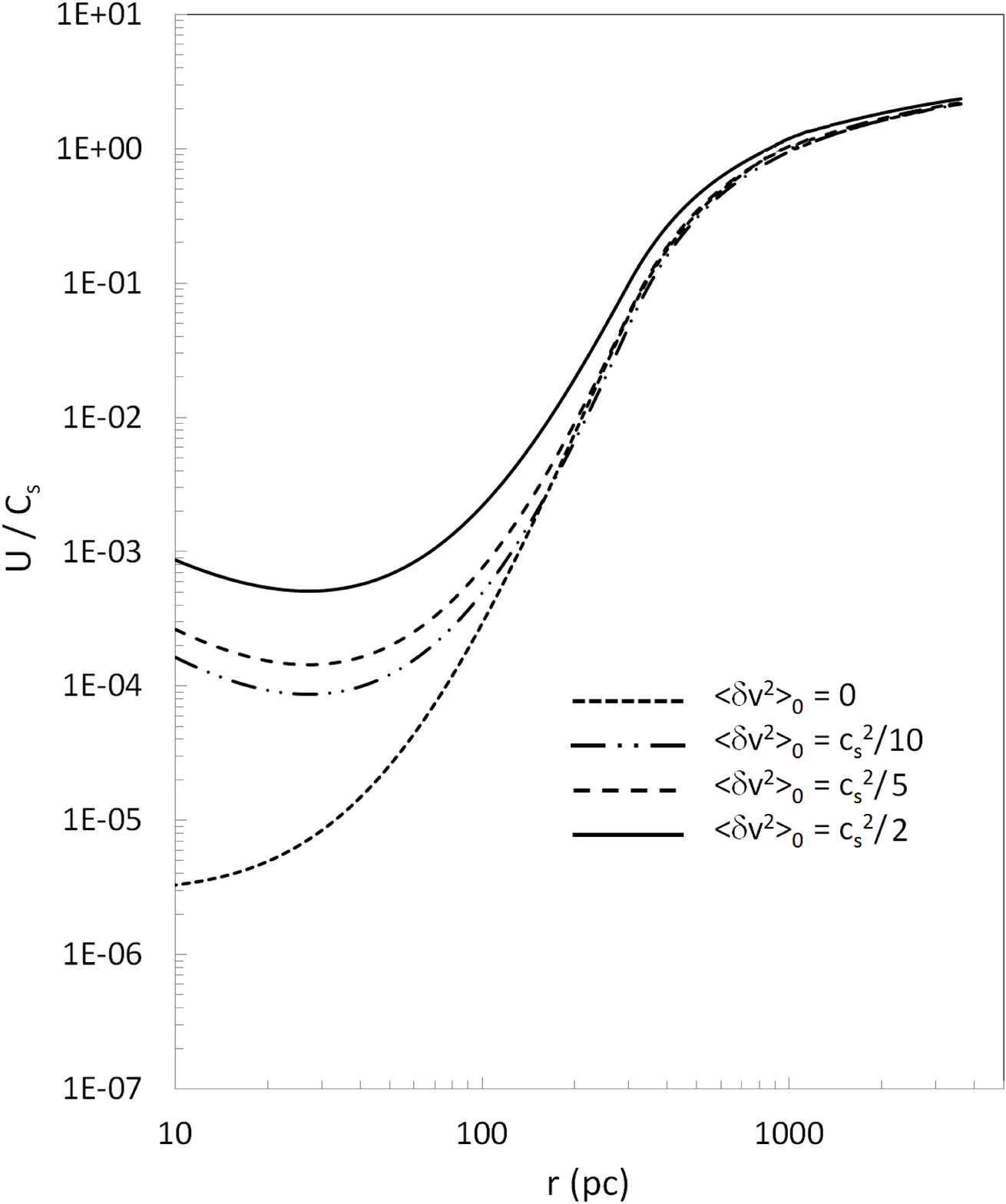}
		\caption {Wind sonic Mach number $\mathcal{M}_s \equiv u/c_s$ as a function of the radius, calculated for models with wave amplitudes $\left< \delta V^2 \right>_0 = 0$ (dotted), $0.1 c_s^2$ (dot-dashed), $0.2 c_s^2$ (dashed) and $0.5 c_s^2$ (solid). We adopted $T=10^5$K, and a NFW profile with $\rho_0 = 10^{-21}$g cm$^{-3}$ (dashed) and $R_S = 200$pc.}
	\label{fig3}
\end{figure}

As expected, the wind mass loss rate increases with the initial energy density of the waves. As shown in Table 1, the mass loss rate is basically proportional to the turbulent Mach number for $\left< \delta V^2 \right>_0/c_s^2>0.1$. A linear regression analysis provide $\dot{M} \propto \left< \delta V^2 \right>_0 ^{0.88}$, with reduced $\chi^2 = 0.96$. The increase in the mass loss rate with the wave energy density is noticeable in both dark matter profiles. For $\left< \delta V^2 \right>_0/c_s^2<0.1$ the role of acoustic waves is irrelevant. 

The extra acceleration of the wind from waves below the first critical point results in increased mass loss rates. In order to continue presenting a physical stationary solution at the second critical position the initial velocity of the obtained solution increases with $\left< \delta V^2 \right>$, which results in increased mass loss rates. In a marginally supersonic regime, i.e. $\left< \delta V^2 \right>^{1/2} \sim c_s$, the mass loss rate is increased by an order of magnitude. On the other hand, the waves play a minor role on the terminal wind velocities. This because the second critical point is located too far from the central region of the galaxy and, at that location, the energy density of the waves is too low. It means that beyond the second critical position, where further acceleration (above a value very close to the sound speed) should occur, the waves have little energy to effectively work over the wind.

\subsubsection{The case of $\epsilon \neq 0$ and finite $L$}

The picture described above changes when we allow the waves to decay. As the waves are damped the radial gradient of the wave energy density becomes steeper. As a consequence, the net force generated by the wave kinetic pressure over the plasma increases. However, a not so obvious result is that this extra pressure actually helps to radially decrease the wind speed if $r < r_{\rm crit,1}$. Within this region, as discussed above, $\partial_r u$ is negative and a short damping length $L$ worsens this scenario (see Eq.10) by reducing $u$ even more. Typically, the wave damping length is short enough for the wave energy to be mostly released at radii smaller than that of the acceleration region, i.e. $L < r_{\rm crit,1}$. At $r > r_{\rm crit,2}$ the wind stationary solution becomes similar to the standard isothermal case, as given in Table 1. 

Actually, our results show that the role of the wave damping depends on the dark matter profile that is considered. For a NFW profile the mass loss rates for the models without waves are very similar to those with damped waves, for $L < 50$pc. If $L > 50$pc the mass loss rates of the models with damped waves become similar to those without damping. This relation is not obtained for the log-profile though, for which the models with and without wave damping present similar mass loss rates and wind terminal velocities. 

Obviously, the model of an isothermal wind with no wave pressure is a crude approximation of a much more complex phenomenon. Nevertheless, this model represents an interesting analytical benchmark for a more complete understanding of the problem of mass loss in dwarf galaxies. The full solution of the thermal wind as given above, i.e. non-isothermal and including the net forces of acoustic waves within the galaxy, are presented in the next section where we compare those results to the simplest model described above.

\subsection{Non-isothermal winds} 

Now let us assume that the temperature gradient is not zero. Unfortunately, it is not possible to consistently describe all the external heating sources that may occur at the early stages of dwarf galaxies. Therefore we will simply assume that $\Gamma$ in Eq.9 is irrelevant compared to the other terms (at least for $T>10^4$K), namely the radiative cooling, wave heating and adiabatic expansion. This is possibly not a good approximation from the time when a large number of stars is already formed because their luminosity may become dominant in the energy budget of the interstellar gas. Eqs. 8 and 9 combined now lead to:

\begin{eqnarray}
\frac{1}{u} \frac{\partial u}{\partial r} \left[ u^2 - \frac{5 c_s^2}{3}-\frac{\left< \delta V^2 \right>}{2} \frac{1+3\mathcal{M}}{1+\mathcal{M}} \right]= -\frac{GM(r)}{r^2}+\frac{2 c_s^2}{r} \nonumber \\ +\frac{4}{3r} + \frac{\left< \delta V^2 \right>}{L}\left[ \frac{1}{2}-\frac{2}{3}\left( \frac{\mathcal{M}+1}{\mathcal{M}} \right) \right] + \frac{2 n \Lambda }{3 \mu m_H u}.
\end{eqnarray}

The cooling function $\Lambda(T)$ used in our calculations was obtained from the interpolation of a 
cooling efficiency table for an optically thin gas for $T \geq 10^4$K (Gnat \& Sternberg 2007), for a low metallicity gas with $Z=10^{-2}$. For temperatures $T < 10^4$K, we assume that ionization heating becomes important in such a way that $\partial_r T \simeq 0$.

We integrated Eq. 18 above, as well as Eq. 9, in order to obtain both the velocity and temperature radial distributions. 
In Figure 4 we present the profiles of $T$ and $u/c_s$ for two models with $T_0 =10^5$K, $L=100$pc, NFW gravitational potential with $\rho_0 =10^{-21}$g cm$^{-3}$ and $R_S = 200$pc.  The difference between the two is the initial energy density of waves that was set as $\left< \delta V^2 \right>_0/c_s^2=10$ (solid lines) and 20 (dashed lines). It is obvious that the cooling and adiabatic expansion initially result in a temperature decrease while, on the other hand, the damping of the wave flux heats the gas. At lower radii the effect of the wave damping is negligible and the temperature gradient is negative for both cases. We found that the radiative cooling will dominate over the adiabatic expansion as main coolant of the plasma if the gas density is larger than $n_0 \sim 3$cm$^{-3}$.

As $r \rightarrow L$, there is an increase of the energy released by the waves for the plasma heating. In both models shown in Fig. 4 the wave heating becomes larger than the cooling processes and an increase in temperature is observed. For $\left< \delta V^2 \right>_0/c_s^2=20$ the temperature at $r \sim 100$pc becomes even larger than the initial setup temperature of $T_0 = 10^5$K. Depending on the initial wave energy density chosen it is possible to obtain temperatures as high as $10^6$K. In both cases though, the temperature at the second critical point is $T \sim 10^4$K. 

Again we found no significant differences in terminal wind velocities, being $u_\infty /c_{s,0} \simeq 2.2$ and $2.4 $ for $\left< \delta V^2 \right>_0/c_s^2=10$ and $20$, respectively. The reason is that most of the wave energy is deposited below the second critical point. However, notice that the Mach number is much different, $\mathcal{M}_\infty \sim 7.5$, because of the lower temperature at larger radii. 

\begin{figure}
	\centering
		\includegraphics[scale=0.12]{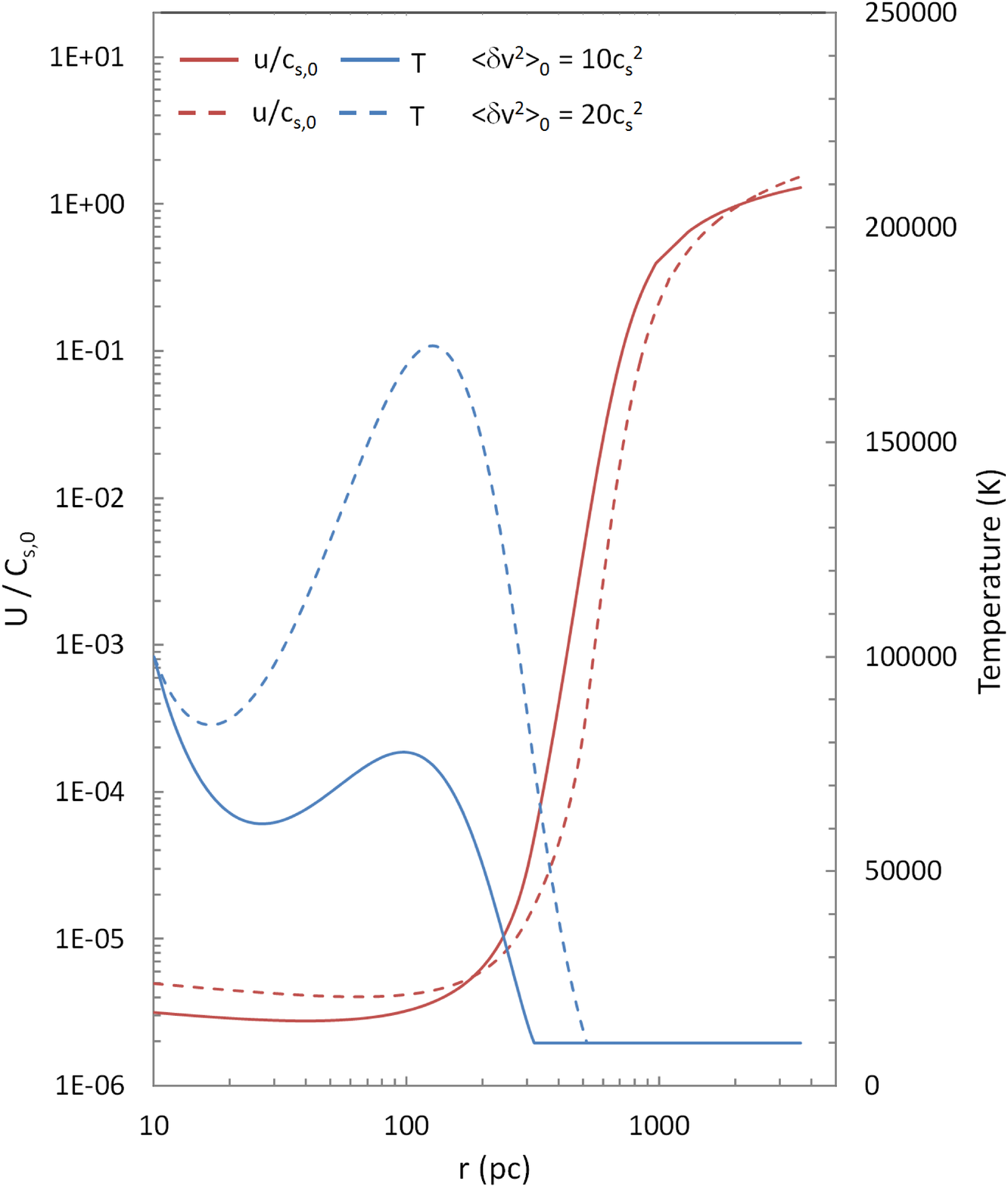}
		\caption {Wind sonic Mach number $u/c_{s,0}$ and gas temperature $T$ as a function of the radius, for $\left< \delta V^2 \right>_0 = 10 c_s^2$ (solid) and $20 c_s^2$ (dashed), both with $T_0=10^5$K and $L = 100$pc.}
	\label{fig4}
\end{figure}

Similar to what is observed in the isothermal cases, the fact that most of the wave energy density is deposited below the second critical position is relevant for the mass loss rate. We obtained $\dot{M} \simeq 1.2$ and $1.7\times 10^{-4}$M$_{\odot}$ yr$^{-1}$ for $\left< \delta V^2 \right>_0/c_s^2=10$ and $20$, respectively. The mass loss rates obtained for the non-isothermal cases are more than 1 order of magnitude smaller than the obtained for isothermal winds with similar physical parameters, with $T_0 = 10^5$K.  The obtained mass loss rates though are similar to the observed mass loss rates of colder isothermal winds ($T \simeq 10^4$K). This because the main acceleration of the non-isothermal winds occurred near to the second critical position, i.e. $r>400$pc, where the temperature is already low, close to $10^4$K. 
We run several models of non-isothermal winds and all presented very low mass loss rates as well. If any relevant external heating source is included though, in the sense that the temperature does not diminishes substantially at $r<r_{\rm crit,2}$, the mass loss rates would happen to be much larger, comparable to the isothermal models with larger temperatures. It is interesting that, despite of its increased complexity, the wind profiles obtained from non-isothermal models are not relevantly different from those obtained with the simplified isothermal set of equations.

\section{Discussion}

The results obtained from the semi-analytical modeling of galactic winds presented in this work show that dwarf galaxies are very likely to present steady winds with significant mass loss rates, but with relatively slow terminal speeds. Even in the case with low wave energy density, i.e. basically pure thermal winds, the mass loss rates reach values as large as $10^{-3} n_{200}^{gas}$M$_{\odot}$yr$^{-1}$, depending on the plasma temperature and the dark matter potential profile. This value can even be an order of magnitude larger if kinetic wave pressure is added at the bottom of the galactic wind. 
Therefore for the case of an initial total gas mass of $M_{\rm gas} \simeq 10^{7}$M$_{\odot}$, as considered in the present work, the galaxy would loose the totality of its gas mass in timescales of order of $10^8 - 10^10$yrs. 
This can be directly compared to the observational estimates. Schwartz \& Martin (2004), for example, detected outflowing gas from six dwarf galaxies with starbursts at very low velocities, typically  $u_\infty \leq 40$km s$^{-1} \sim c_s$, as predicted from our models. They also infered the outflow gas masses as $M_{\rm outflow} \simeq 10^{6} - 10^{10}$M$_{\odot}$, also in agreement with the total gas mass that such a system could loose during its evolution based on our model. Large mass losses from dSph have also been estimated for the Local Group objects based on chemical evolution models (Kirby et al. 2011). Slow galactic winds for dwarf galaxies have also been  been detected by other studies, e.g. van Eymeren et al. (2010).

Obviously, the mass loss rate is increased once wave pressure is added at small radii. Kinetic wave pressure is typically related to SNe explosions. Few explosions of the massive stars from young stellar populations would result in a turbulent interstellar gas, or even in the direct blow of the ejecta out of the galaxy. 
Supernovae (SNe) are possibly the most important source of turbulence in the ISM of dwarf galaxies, which have been generically considered in this work as a preset energy density of acoustic waves. Let us consider Type II SNe as the most important energy sources in an early dwarf galaxy. In a stationary case the wave energy flux equals the total power released from the stellar explosions into kinetic motions:

\begin{eqnarray}
\dot{E}_{\rm wav}\simeq \xi \dot{E}_{\rm SN} \nonumber \\
\int \frac{ \rho \left< \delta V^2 \right> u_\infty}{L} dV \simeq \xi \mathcal{R}_{\rm SN} \varepsilon_{\rm SN}, 
\end{eqnarray}

\noindent
being $\mathcal{R}_{\rm SN}$ the supernova rate, $\varepsilon_{\rm SN}\simeq 10^{51}$erg the total energy released by the SN, and $\xi$ the fraction of the total energy converted into kinetic motions of the interstellar medium. 

Notice that for $\xi = 0.01$ and $n_0 = 10^3$cm$^{-3}$ it is possible to obtain $\left< \delta V^2 \right> \sim 10 c_s^2$ with $\mathcal{R}_{\rm SN} \sim 10^{-7}$SN yr$^{-1}$. This is a very low SNe explosion rate, compared to the estimates for typical early evolutionary stages of dSph ($\mathcal{R}_{\rm SN} \sim 10^{-6}$SN yr$^{-1}$). Therefore, it is reasonable to assume the existence of a continuous and isotropic acoustic wave pressure, from supersonic turbulence, at the central regions of the galaxy. 

Many models have been proposed in the literature to account for the mass loss rates of SNe driven winds using simplified prescriptions (e.g. Holzer \& Axford 1970, Larson 1974, Bradamante et al. 1998, Lanfranchi \& Matteucci 2007, C\^ot\'e et al. 2012, Sharma \& Nath 2012, and others), but none have presented a systematic study of quasi-stationary solutions as has been presented in this work. 

Compared to previous works, an important difference from our results is the dependence of the wind mass loss rate with the profile of dark matter. This result shows that simply comparing the volumetric integrations of SNe energy/mass feedback and gravitational binding energy may not the best relation to determine if the wind will be launched or not. This because the radial dependence of the acceleration of gravity is strongly dependent on the dark matter profile. Actually, this parameter is the one to be compared to the gradient of the energy density released by SNe. Another basic result, which gives different predictions compared to previous works, is the possibility of stationary winds even in the presence of very low star formation rate. Large mass loss rates are also obtained from pure thermally driven winds, given that any internal or external source keeps the interstellar gas heated at minimum temperature $T>10^4 - 10^5$K. Therefore, the quenching of star formation may occur at very early stages of the galaxy evolution. Estimates predict a reionization of early Universe at $z \sim 5 - 11$ (e.g. Robertson et al. 2010) due to the formation of the first stars, which corresponds to a cosmic age of $400 - 1100$Myrs in a standard $\Lambda$CDM Universe. If the heating of the galactic ISM occurs at these times, and considering the timescales for mass loss from our models, we may estimate a cut-off in the star formation at $t \sim 1.5$Gyrs.  

Other key timescales for the galaxy evolution, such as the star formation timescale and the gas infall timescale, also play a major role on the time evolution of the equilibrium gas fraction since both timescales  evolve strongly with redshift (see Khochfar \& Silk 2011).
The accretion rate related to the cold gas infall is often related to the dark matter potential as $\dot{M}_{\rm acc} \sim f_b \dot{M}_{\rm DM}$, being $f_b \sim 0.165$ the baryon mass fraction (Dekel et al. 2009). For the dwarf galaxies observed at $z \rightarrow 0$ it is expected that their dark matter haloes did not change their masses much since their formation, except for merging processes. Larger galaxies, on the other hand, would evolve for longer times with an approximate growth rate given by $\dot{M}_{\rm DM}(z) \sim 0.033 f_b^{-1} M_{10}^{1.15} (1+z)^{2.25}$M$_{\odot}$yr$^{-1}$ (Neistein et al. 2006), where $M_{10}$ represents the dark matter halo mass in units of $10^10$ solar masses. At $z = 9$, for a $10^9$ solar masses dark matter halo, one then obtain $\tau_{\rm acc} (z=9) \sim f_b M_{\rm DM} \dot{M}_{\rm acc}^{-1} \sim 3 \times 10^7$yr. The accretion is then supposed to occur much faster than the gas loss timescales given above. The typical timescales for star formation, on the other hand, is estimated as $\tau_{\rm SF} \sim f_b M_{\rm DM} SFR^{-1}$, being $SFR$ the star formation rate. At $z = 9$, for a $10^9$ solar masses dark matter halo, the $SFR \sim 10^{-3} - 10^{-2}$M$_{\odot}$yr$^{-1}$ (see Weisz et al. 2011), which leads to $\tau_{\rm SF} \geq 10^{10}$yrs. The two later timescales show that the intergalactic baryons must fall into the dwarf dark matter haloes, but the timescales needed for complete conversion into stars are much larger than the typical mass loss timescale. As a result, only few percent of the gas mass forms stars, and most of it is ejected out of the galaxy.
Cases of galaxies with star formation bursts at $z \rightarrow 0$ could be explained by later infall of gas, subsequent to a long period of inactivity.

\subsection{Dark matter and the time dependence of mass loss rates in dSphs} 

Numerical simulations of the dynamical evolution of the matter in the early Universe predict the formation of a large number of low mass dark matter structures, which are supposed to host the dwarf galaxies. These simulations also predict a cusped dark matter profile in the center of these objects (e.g. Navarro, Frenk \& White 1996). However, recent observational studies have pointed to less steep (cored) distributions instead (Burkert 1995; van den Bosch et al. 2000; de Blok \& Bosma 2002; Kleyna et al. 2003; Simon et al. 2005; Walker et al. 2009; Governato et al. 2010; Oh et al. 2011; Del Popolo 2012; Jardel \& Gebhardt 2012). 

The discrepancy between the cusped and core profiles is possibly related to the evolution of these galaxies. Since these low mass structures are typically satellites of massive galaxies, models predict that the interaction between their gravitational potentials stirs up the initially central-cusped dark matter distribution, resulting in cored profiles (Mayer et al. 2001a, 2001b; Klimentowski et al. 2007, 2009; Kazantzidis et al. 2011; Lokas, Karantzidis \& Mayer 2012). Also, at later evolutionary stages of these galaxies the baryonic matter may interfere in the dynamics of the dark matter. Possibly, stellar dynamics also result in core-like dark matter density profiles (Governato et al. 2010; de Souza et al. 2011; Pontzen \& Governato 2012; Governato et al. 2012). 

The timescales for changes in the dark matter distribution to occur range from  $\tau_{\rm ch} \sim$ hundreds of Myrs to Gyrs. These can be directly compared to the timescale for the galactic wind of a NFW profile to remove most of the interstellar gas $\tau_{\rm wind} \sim M_{\rm gas}/\dot{M_{\rm NFW}}$. If $\tau_{\rm ch} > \tau_{\rm wind}$, the sharp dark matter potential would drive a faster mass loss process that removes most the gas from the galaxy, and star formation is quenched until another gas infall from the intergalactic medium occurs (infall regulated star formation events). On the other hand if $\tau_{\rm ch} < \tau_{\rm wind}$, the gravitational potential evolves fast enough for the galactic mass loss to slow down. The interstellar gas would be only partially removed, and star formation would occur continuously.

\section{Conclusions}

Galactic winds play an important role on the mass feedback of the intergalactic medium, particularly on its chemical evolution, as well as in the galactic star formation history. Dwarf galaxies are known to present relatively more massive dark matter haloes, and systematically lower metallicities, compared to the more massive galaxies. The presence of strong galactic winds during the early stages of their evolution could be responsible for quenching star formation and help reduce the baryon fraction of their total masses. Theoretical modeling of dwarf galaxy winds typically considers supernovae (SNe) as main, or even only, sources of energy and momentum. 

In this work we apply a Parker-like wind model to the physical properties of dwarf galaxies in order to study the role of the thermal pressure on their wind acceleration. We studied different dark matter potentials, temperature distributions and SNe driven turbulence. Despite of the general picture of chaotic and stochastic galactic winds (e.g. SNe driven winds) it is possible to drive quasi-stationary thermally driven flows. Also, we showed that the stationary solution presents two critical points, which will fully determine the wind mass loss rates and wind terminal velocities.

As main results, we found that:

\begin{itemize}

\item  physical quasi-stationary solutions of winds in dwarf galaxies are possible if the interstellar gas is hot ($T>10^4$K), depending on the gravitational potential;

\item  dwarf galaxies are likely to present galactic winds at very early evolutionary stages, even when presenting very low star formation, since thermally driven winds showed to be quite efficient;

\item  cusped dark matter profiles, e.g. the Navarro-Frenk-White profile, result in larger mass loss rates, compared to cored or less steep profiles.

\end{itemize}

The models also showed that increased turbulent motions at the bottom of the winds, possibly associated to star formation activity, are responsible for an increase in mass loss rate. This result is in agreement to the simplified analytical models that compare the binding gravitational energy to the SNe energy thermalized at the ISM, and with numerical simulations of SNe driven galactic winds. However, the wind terminal velocities obtained showed that this parameter is insensitive to the SF activity.

The models described in this work may serve as benchmarks for future numerical simulations, as well as basic wind mass loss rates and terminal velocity prescriptions for chemical and dynamical models of dwarf galaxies that use galactic outflows as a model input. As an example, closed box chemical evolution models include infall and mass loss rates in order to study the metal enrichment of the galaxies taking into account losses of metals to the IGM due to galactic winds. Typically, this has been done by simple energy budget arguments. These models would greatly benefit from the semi-analytical predictions of this work. Also, full hydrodynamical numerical simulations could also be set in order to check the validity of a quasi-steady solution for the wave/thermally driven galactic wind. As basic parameters, the set of a dark matter potential, initial equilibrium distributions for gas temperature and density, and the sources for kinetic waves and heating, would be enough in order to obtain the time evolution of the gas flow, and to compare with our results.

\section*{Acknowledgments}

DFG thanks CNPq (no. 300382/2008-1) and FAPESP (no. 2011/12909-8) for financial support, and the anonymous referee that helped in improving this paper.


\begin{thebibliography}{99}

\bibitem[]{} Armandroff, T. E., Olszewski, E, W. \& Pryor, C. 1995, AJ, 110, 2131

\bibitem[]{} Bradamante, F., Matteucci, F. \& D`Ercole, A. 1998, A\&A, 337, 338

\bibitem[]{} Bullock, J. S. \& Johnston, K. V. 2005, ApJ, 635, 931

\bibitem[]{} Burkert, A. 1995, ApJ, 447, L25

\bibitem[]{} Chen, Y.-M., Tremonti, C. A., Heckman, T. M., Kauffmann, G., Weiner, B. J., Brinchmann, J., Wang, J. 2010, AJ, 140, 445

\bibitem[]{} Chevalier,R. A. \& Clegg,A. W., 1985, Nature, 317, 44

\bibitem[]{} C\^ot\'e, B., Martel, H., Drissen, L., \& Robert, C. 2012, MNRAS, 421, 847

\bibitem[]{} de Blok, W. J. G., \& Bosma, A. 2002, A\&A, 385, 816

\bibitem[]{} de Souza R. S., Rodrigues L. F. S., Ishida E. E. O.,  \& Opher R., 2011, MNRAS, 415, 2969

\bibitem[]{} Dekel A., \& Silk J., 1986, ApJ, 303, 39

\bibitem[]{} Dekel, A., et al. 2009, Nature, 457, 451

\bibitem[]{} Del Popolo, A. 2012, MNRAS, 419, 971

\bibitem[]{} Erb, D. K., Quider, A. M., Henry, A. L. \& Martin, C. L. 2012, ApJ, 759, 26

\bibitem[]{} Fragile, P. C., Murray, S. D., Anninos, P. \& Lin, D. N. C. 2003, ApJ, 590, 778

\bibitem[]{} Fujita, A.; MacLow, M. M.; Ferrara, A. \& Meiksin, A. 2004, ApJ, 613, 159

\bibitem[]{} Governato, F., Brook, C., Mayer, L., Brooks, A., Rhee, G., Wadsley, J., Jonsson, P., Willman, B., Stinson, G., Quinn, T., \& Madau, P. 2010, Nature, 463, 203

\bibitem[]{} Governato, F., Zolotov, A., Pontzen, A., Christensen, C., Oh, S. H., Brooks, A. M., Quinn, T., Shen, S., \& Wadsley, J. 2012, MNRAS, 422, 1231

\bibitem[]{} Gnat, O., \& Sternberg, A. 2007, ApJ, 168, 213

\bibitem[]{} Jardel, J. R. \& Gebhardt, K. 2012, ApJ, 746, 89

\bibitem[]{} Kazantzidis, S., Lokas, E. L., Callegari, S., Mayer, L., \& Moustakas, L. A. 2011, ApJ, 740, 24

\bibitem[]{} Kirby, E. N., Martin, C. L. \& Finlator, K. 2011, ApJ, 742, L25

\bibitem[]{} Kirby, E. N., Lanfranchi, G. A., Simon, J. D., Cohen, J. G., Guhathakurta, P. 2011, ApJ, 727, 78

\bibitem[]{} Kleyna, J. T., Wilkinson, M. I., Gilmore, G., \& Evans, N. W. 2003, ApJ, 588, L21

\bibitem[]{} Klimentowski, J., Lokas, E. L., Kazantzidis, S., Prada, F.,Mayer, L., \& Mamon, G. A. 2007, MNRAS, 378, 353

\bibitem[]{} Klimentowski, J., Lokas, E. L., Kazantzidis, S., Mayer, L., \& Mamon, G. A. 2009, MNRAS, 397, 2015

\bibitem[]{} Khochfar, S. \& Silk, J. 2011, MNRAS, 410, 42

\bibitem[]{} Lanfranchi, G. A., \& Matteucci, F. 2007, A\&A, 468, 927

\bibitem[]{} Larson, R. B. 1974, MNRAS, 169, 229

\bibitem[]{} Lokas, E. L., Kazantzidis, S., \& Mayer, L. 2012, ApJL, 751, 15

\bibitem[]{} Mac Low, M.-M., \& Ferrara, A. 1999, ApJ, 513, 142

\bibitem[]{} Martin, C. L., Shapley, A. E., Coil, A. L., Kornei, K. A., Bundy, K., Weiner, B. J., Noeske, K. G.\& Schiminovich, D. 2012, ApJ, 760, 127

\bibitem[]{} Mateo, M. L. 1998, ARA\&A, 36, 435

\bibitem[]{} Mayer, L., Governato, F., Colpi, M., Moore, B., Quinn, T., Wadsley, J., Stadel, J., \& Lake, G. 2001a, ApJ, 559, 754

\bibitem[]{} Mayer, L., Governato, F., Colpi, M., Moore, B., Quinn, T., Wadsley, J., Stadel, J., \& Lake, G. 2001b, ApJL, 547, 123

\bibitem[]{} Mori, M., Ferrara, A., \& Madau, P., 2002, ApJ, 571, 40

\bibitem[]{} Navarro, J. F., Frenk, C. \& White, S. D. M. 1996, ApJ, 462, 563

\bibitem[]{} Neistein, E., van den Bosch, F. C., \& Dekel, A. 2006, MNRAS, 372, 933

\bibitem[]{} Oh, S., Brook, C., Governato, F., Brinks, E., Mayer, L., de Blok, W. J. G., Brooks, A., \& Walter, F. 2011, AJ, 142, 24

\bibitem[]{} Peeples M. S., \& Shankar F., 2011, MNRAS, 417, 1387

\bibitem[]{} Persic, M., Salucci, P., \& Stel, F. 1996, MNRAS, 281, 27

\bibitem[]{} Pontzen, A., \& Governato, F. 2012, MNRAS, 421, 3464

\bibitem[]{} Robertson, B., Bullock, J. S., Font, A. S., Johnston, K. V. \& Hernquist, L. 2005, ApJ, 632, 872

\bibitem[]{} Robertson, B. E., Ellis, R. S., Dunlop, J. S., McLure, R. J. \& Stark, D. P. 2010, Nature, 468, 49

\bibitem[]{} Ruiz, L. O., Falceta-Gon\c calves, D., Lanfranchi, G. A. \& Caproni, A. 2013, MNRAS, 429, 1437

\bibitem[]{} Salucci, P. et al. 2012, MNRAS, 420, 2034

\bibitem[]{} Schwartz, C. M. \& Martin, C. L. 2004, ApJ, 610, 201

\bibitem[]{} Sharma, M. \& Nath, B. B. 2012, arXiv:1209.0242

\bibitem[]{} Silich, S., Tenorio-Tagle, G., \& Mu\~noz-Tu\~n\'on, C., 2003, ApJ, 590, 791

\bibitem[]{} Simon, J. D., Bolatto, A. D., Leroy, A., Blitz, L. \& Gates, E. L. 2005, ApJ, 621, 757

\bibitem[]{} Somerville, R. S. \& Primack, J. R.  1999, MNRAS, 310, 1087

\bibitem[]{} Steidel, C. C., Erb, D. K., Shapley, A. E., Pettini, M., Reddy, N., Bogosavljevic, M., Rudie, G. C., Rakic, O. 2010, ApJ, 717, 289

\bibitem[]{} Stringer, M. J., Bower, R. G., Cole, S., Frenk, C. S. \& Theuns, T. 2012, MNRAS, 423, 1596

\bibitem[]{} Tremonti C. A. et al., 2004, ApJ, 613, 898

\bibitem[]{} van den Bosch, F.C., Robertson, B.E., Dalcanton, J. J., \& de Blok, W.J.G. 2000, AJ, 119, 1579

\bibitem[]{} van Eymeren, J., Koribalski, B. S., López-Sánchez, A. R., Dettmar, R.-J., Bomans, D. J. 2010, MNRAS, 407, 113

\bibitem[]{} Veilleux, S., Cecil, G., \& Bland-Hawthorn, J. 2005, ARA\&A, 43, 769

\bibitem[]{} Walker, M. G., Mateo, M., Olszewski, E. W., Pe\~narrubia, J., Wyn Evans, N., \& Gilmore, G. 2009, ApJ, 704, 1274

\bibitem[]{} Weisz, D. R. et al. 2011, ApJ, 739, 5




\end{thebibliography}
\end{document}